**Mueller matrices for anisotropic metamaterials generated using 4×4 matrix formalism**

P. D. Rogers[1], T. D. Kang[1], T. Zhou[1], M. Kotelyanskii[1,2], and A. A. Sirenko[1]

[1]Department of Physics, New Jersey Institute of Technology, Newark, New Jersey 07102

[2]Rudolph Technologies Inc., Flanders, New Jersey 07836

**Abstract**

Forward models for the Mueller Matrix (MM) components of materials with relative magnetic permeability tensor $\mu \neq 1$ are studied. $4 \times 4$ matrix formalism is employed to produce general solutions for the complex reflection coefficients and MMs of dielectric-magnetic materials having arbitrary crystal symmetry and arbitrary laboratory orientation. For certain orientations of materials with simultaneously diagonalizable $\varepsilon$ and $\mu$ tensors (with coincident principal axes), analytic solutions to the Berreman equation are available. For the single layer thin film configuration of these materials, analytic formulas for the complex reflection and transmission coefficients are derived for orthorhombic or higher crystal symmetry. The separation of the magnetic and dielectric contributions to the optical properties of a material are demonstrated using measurements of the MM at varying angles of incidence.

## 1. Introduction

The calculation of a forward model for the Mueller Matrix (MM) components of a dielectric-magnetic material is critical to the analysis of experimental data obtained from full MM spectroscopic ellipsometry. Through an iterative numerical comparison of the forward model against experimental data, the optical properties of a dielectric-magnetic material can be analyzed. Specifically, models for the relative dielectric permittivity tensor $\varepsilon$ and the relative magnetic permeability tensor $\mu$ can be developed. $4 \times 4$ matrix formalism [1] provides a powerful method to calculate the complex reflection coefficients and the MMs of dielectric-magnetic materials having both arbitrary crystal symmetry and magnetic permeability tensor $\mu \neq 1$. For materials with certain laboratory orientations, simultaneously diagonalizable $\varepsilon$ and $\mu$ tensors (with coincident principal axes), and orthorhombic crystal symmetry or higher, exact solutions for allowed electromagnetic wave propagation in a dielectric-magnetic medium are produced. Analytic solutions for the complex reflection coefficients for $p$ and $s$ polarization states in both semi-infinite and thin film configurations are calculated. For a non-depolarizing medium, forward MM models are determined directly from the complex reflection coefficients of the material. In this paper, forward MM models that match the symmetry of planar metamaterials are calculated by treating their behavior as a continuous anisotropic thin film. We focus on recently published studies pertaining to artificially created planar metamaterials which use modified Lorentzian oscillator models for the diagonal components of the $\varepsilon$ and $\mu$ tensors [2]. Simulations for MM components in the frequency range close to the electric and magnetic resonances are presented. In addition, methods to separate the electric and magnetic effects on the optical properties of a dielectric-magnetic material using MM measurements at oblique angles of incidence (AOI) are illustrated.

## 2. 4×4 matrix formalism

Berreman's $4\times 4$ matrix formalism can accommodate materials with magnetic permeability tensor $\mu \neq 1$ [1]. The Berreman equation describing electromagnetic wave propagation in a crystal is:

$$\frac{d\Psi}{dz} = i\frac{\omega}{c}\Delta\Psi \qquad (1)$$

where $\Psi$ is a an array of the transverse components of the electromagnetic wave $[E_x, H_y, E_y, -H_x]^T$ in the medium. Figure 1 illustrates the refraction of light incident in the $x-z$ plane propagating forward in an anisotropic dielectric-magnetic material. For a crystal with orthorhombic symmetry having principal axes parallel to the $x$, $y$ and $z$ coordinate axes, $\Delta$ in Eq. (1) is a $4\times 4$ matrix [1]:

$$\Delta = \begin{pmatrix} 0 & \mu_{yy} - \dfrac{N_0^2 \sin(\theta_0)^2}{\varepsilon_{zz}} & 0 & 0 \\ \varepsilon_{xx} & 0 & 0 & 0 \\ 0 & 0 & 0 & \mu_{xx} \\ 0 & 0 & \varepsilon_{yy} - \dfrac{N_0^2 \sin(\theta_0)^2}{\mu_{zz}} & 0 \end{pmatrix} \qquad (2)$$

Inserting Eq. (2) into Eq. (1) returns four exact solutions of the form $\psi_l(z) = \psi_l(0)e^{iq_l z}$ with $l = 1, 2, 3$ or $4$, two for each of the $p$ and $s$ polarization states. $p(s)$ refers to radiation parallel (perpendicular) to the plane of incidence. $q_{zp}$ and $q_{zs}$ are the eigenvalues associated with $p$ and $s$ polarizations, respectively and constitute the $z$ components of the wave vectors in the medium. These are:

$$q_{zp} = \frac{\omega}{c}\sqrt{\varepsilon_{xx}}\sqrt{\mu_{yy} - \frac{N_0^2 \sin^2(\theta_0)}{\varepsilon_{zz}}} \qquad (3)$$

$$q_{zs} = \frac{\omega}{c}\sqrt{\mu_{xx}}\sqrt{\varepsilon_{yy} - \frac{N_0^2 \sin^2(\theta_0)}{\mu_{zz}}} \qquad (4)$$

Figure 1 shows $q_{zp}$ and $q_{zs}$. The $x$ component of the wave vector is constant for all of the incident and refracted waves. It is through these equations (eigenvalues of the Berreman equation) that information about the anisotropic optical properties of the medium [3] enters into the calculation of the complex reflection coefficients and, in turn, MM elements. For example, the anisotropic $\varepsilon$ and $\mu$ tensors and the consequent differences between $q_{zp}$ and $q_{zs}$ are responsible for the two refracted waves shown in Figure 1.

## 3. Analytic Formulas

One of the key benefits of using $4\times 4$ matrix formalism to calculate complex reflection and transmission coefficients is that procedures for matching electromagnetic boundary conditions are automatically built in to the method when both incident and, in the case of thin films, substrate media are isotropic and non-magnetic. For each polarization state there are two eigenvectors representing forward and backward propagating waves. In $4\times 4$ matrix formalism, the complex reflection coefficients $r_{pp}(\omega)$ and $r_{ss}(\omega)$ and the complex transmission coefficients $t_{pp}(\omega)$ and $t_{ss}(\omega)$ are calculated from the eigenvectors of Eq. (1) via the solution of simultaneous boundary value equations relating to the continuity of the electric and magnetic fields at the media interface(s). For semi-infinite samples, backward propagating waves are not considered. For thin film samples, retention of the two backward propagating waves is essential to the proper calculation of the complex reflection and transmission coefficients as well MM elements.

*3.1 Semi-Infinite Sample*

For a semi-infinite material, the two eigenvectors representing the forward propagating waves are used to calculate the complex reflection coefficients for *p* and *s* polarized radiation. The complex reflection coefficients are:

$$r_{pp} = \frac{\varepsilon_{xx} k_{z0} - N_0^2 q_{zp}}{\varepsilon_{xx} k_{z0} + N_0^2 q_{zp}} \qquad (5)$$

$$r_{ss} = \frac{\mu_{xx} k_{z0} - q_{zs}}{\mu_{xx} k_{z0} + q_{zs}}. \qquad (6)$$

In Eq. (5) and Eq. (6), the complex reflection coefficients are expressed as functions of the *z* components of the incident and refracted wave vectors which themselves take into account the anisotropic characteristics of the medium. Complex reflection coefficients stated in this formalism have been used in the study of media with indefinite permittivity and permeability tensors [4]. This formalism also allows for immediate analysis of the intriguing property of impedance matching. Consider an isotropic medium. From Eq. (5), at normal incidence, $r_{pp}$ is zero when $N_0 = \sqrt{\varepsilon/\mu}$. A similar result can be obtained for $s$ polarization from Eq. (6). These relationships are known as the impedance matching condition. It provides the condition for zero reflection at normal incidence even though the indices of refraction of the incident medium $(N_0)$ and the index of refraction of the material ($\sqrt{\varepsilon\mu}$) are completely different. With incidence from vacuum, this condition is satisfied if $\varepsilon = \mu$. Clearly, this is only possible if the material is magnetic and provides confirmation that the material has magnetic permeability $\mu \neq 1$. In practice, it is difficult to achieve impedance matching because both the real and imaginary parts of the dielectric and magnetic tensors must be identical. Evidence of impedance matching in metamaterials was found by Grigorenko et al. in 2005 [5].

*3.2 Thin Film Sample*

For a single layer thin film material, all four eigenvectors and eigenvalues are used in the calculation of both the complex reflection and transmission coefficients. Both incident and substrate media are assumed to be isotropic, non-magnetic materials. The $z$ components of the incident and substrate wave vectors are $k_{z0} = \frac{\omega}{c} N_0 \cos(\theta_0)$ and $k_{z2} = \frac{\omega}{c} N_2 \cos(\theta_2)$, respectively. The dielectric-magnetic thin film has thickness $d$ and is described by $\varepsilon$ and $\mu$ tensors each having orthorhombic symmetry. We assume that the $\varepsilon$ and $\mu$ tensors are simultaneously diagonalized and have coincident principal axes. Higher symmetries can easily be derived from the orthorhombic case. The crystal is aligned such that its principal axes are coincident with the laboratory axes. Light is again incident in the $x - z$ plane (see Figure 1). Using $4 \times 4$ matrix formalism which matches the appropriate boundary conditions for the electric and magnetic field vectors obtained from the thin film eigenvectors, we were able to derive analytic expressions for both $p$ and $s$ polarizations.

The complex reflection and transmission coefficients for *p* polarized radiation are:

$$r_{pp} = \frac{q_{zp}\cos(q_{zp}d)\left(\frac{N_2}{N_0}k_{z0} - \frac{N_0}{N_2}k_{z2}\right) + i\left(\frac{N_0 N_2 q_{zp}^2}{\varepsilon_{xx}} - \frac{\varepsilon_{xx}k_{z0}k_{z2}}{N_0 N_2}\right)\sin(q_{zp}d)}{q_{zp}\cos(q_{zp}d)\left(\frac{N_2}{N_0}k_{z0} + \frac{N_0}{N_2}k_{z2}\right) - i\left(\frac{N_0 N_2 q_{zp}^2}{\varepsilon_{xx}} + \frac{\varepsilon_{xx}k_{z0}k_{z2}}{N_0 N_2}\right)\sin(q_{zp}d)}$$

$$t_{pp} = \frac{2k_{z0}q_{zp}}{q_{zp}\cos(q_{zp}d)\left(\frac{N_2}{N_0}k_{z0} + \frac{N_0}{N_2}k_{z2}\right) - i\left(\frac{N_0 N_2 q_{zp}^2}{\varepsilon_{xx}} + \frac{\varepsilon_{xx}k_{z0}k_{z2}}{N_0 N_2}\right)\sin(q_{zp}d)}$$

. (7)

The complex reflection and transmission coefficients for *s* polarized radiation are:

$$r_{ss} = \frac{q_{zs}\cos(q_{zs}d)(k_{z0} - k_{z2}) + i\left(\frac{q_{zs}^2}{\mu_{xx}} - k_{z0}k_{z2}\mu_{xx}\right)\sin(q_{zs}d)}{q_{zs}\cos(q_{zs}d)(k_{z0} + k_{z2}) - i\left(\frac{q_{zs}^2}{\mu_{xx}} + k_{z0}k_{z2}\mu_{xx}\right)\sin(q_{zs}d)}$$

$$t_{ss} = \frac{2k_{z0}q_{zs}}{q_{zs}\cos(q_{zs}d)(k_{z0} + k_{z2}) - i\left(\frac{q_{zs}^2}{\mu_{xx}} - k_{z0}k_{z2}\mu_{xx}\right)\sin(q_{zs}d)}$$

. (8)

The formulas are functions of the optical properties of the film material as well as the characteristics of both incident and substrate media. For example, in a vacuum-thin film-vacuum configuration, the first terms in the numerator of each of the complex reflection coefficients become zero. This simpler form is applicable to many experimental configurations and will be used in the analysis of planar metamaterials below.

In order to verify the accuracy of our analytical expressions, we have calculated the complex reflection and transmission coefficients for the cases of the semi-infinite sample, and a single layer film on a semi-infinite substrate using both our numerical implementation of the $4\times 4$ matrix algorithm and the analytical expressions in Eq. (5) through Eq. (8). We found that the results coincide within the rounding errors of the $4\times 4$ matrix algorithm. This analysis was performed for a variety of conditions including negative permittivity and permeability values, which are expected to be observed in metamaterials.

## 4. Mueller matrices of a planar metamaterial

Four complex reflection coefficients constitute the $2\times 2$ Jones matrix of the medium:

$$\begin{pmatrix} r_{pp}(\omega) & r_{ps}(\omega) \\ r_{sp}(\omega) & r_{ss}(\omega) \end{pmatrix}. \qquad (9)$$

For the sample symmetry and the experimental configurations assumed in this paper, the off diagonal elements of the Jones matrix are zero. For experimental data, intensity measurements are required and the 16 element MM is utilized for this purpose. For non-depolarizing materials, there are well established formulas to transform the Jones matrix to a full MM [3] and Eq. (10) is the transformation formula applicable when the off diagonal Jones matrix elements are both zero.

$$\begin{pmatrix} \frac{1}{2}\left(|r_{pp}|^2+|r_{ss}|^2\right) & \frac{1}{2}\left(|r_{pp}|^2-|r_{ss}|^2\right) & 0 & 0 \\ \frac{1}{2}\left(|r_{pp}|^2-|r_{ss}|^2\right) & \frac{1}{2}\left(|r_{pp}|^2+|r_{ss}|^2\right) & 0 & 0 \\ 0 & 0 & \Re\left(r_{pp}r_{ss}^*\right) & \Im\left(r_{pp}r_{ss}^*\right) \\ 0 & 0 & -\Im\left(r_{pp}r_{ss}^*\right) & \Re\left(r_{pp}r_{ss}^*\right) \end{pmatrix} \quad (10)$$

In summary, the MM of a dielectric-magnetic material is produced from its complex reflection coefficients which are, in turn, calculated from its frequency dependent $\varepsilon$ and $\mu$ tensors. Accordingly, to produce a MM, accurate complex reflection formulas appropriate to the orientation of the crystal must be available. In addition, models for the dielectric and magnetic functions of the material are required for input into these reflection formulas. Eq. (10) illustrates that, for our configuration, there will be eight non-zero MM elements. However, only four of these terms are independent. Procedures for calculating the forward model of a MM for a planar metamaterial will now be discussed.

The study of metamaterials has been of interest since the late 1960's when V. G. Veselago first explored the properties of isotropic materials having simultaneous negative $\varepsilon$ and $\mu$ tensors [6]. To date, there have been relatively few spectroscopic studies of metamaterials which analyze their reflection properties using oblique angles of incidence. Driscoll *et al.* have done one such study using a planar array of split-ring resonators (SRRs) [2]. Reflection and transmission intensities were recorded for the single *s* polarization at varying angles of incidence. These results were fitted using the Fresnel equations to model the optical properties of the metamaterial as though it behaved as a continuous anisotropic thin film crystal.

These results are important to our study of MMs because the frequency dependent models of the material's $\varepsilon$ and $\mu$ tensors together with our Eq. (7) and Eq. (8) enable the calculation of predictive MMs of this planar metamaterial. In the Driscoll experimental configuration, the $\varepsilon$ and $\mu$ tensors have the following anisotropic symmetry:

$$\varepsilon(\omega) = \begin{pmatrix} \varepsilon_{xx}(\omega) & 0 & 0 \\ 0 & \varepsilon_{yy}(\omega) & 0 \\ 0 & 0 & 1 \end{pmatrix}$$

$$\mu(\omega) = \begin{pmatrix} 1 & 0 & 0 \\ 0 & 1 & 0 \\ 0 & 0 & \mu_{zz}(\omega) \end{pmatrix} \quad (11)$$

The tensors are described by the modified Lorentzian oscillator models given in Eq. (12). Fitted parameters for these models are given in Table 1. The $\varepsilon_{yy}(\omega)$ response was not analyzed in the Driscoll paper.

$$\varepsilon_{xx}(\omega) = \varepsilon_s - \frac{A_e \omega_p^2}{\omega^2 - \omega_{e0}^2 + i\omega\gamma_e}$$

$$\mu_{zz}(\omega) = 1 - \frac{A_m \omega^2}{\omega^2 - \omega_{m0}^2 + i\omega\gamma_m} \quad (12)$$

The general formulas for thin films derived using $4 \times 4$ matrix formalism are used to calculate the complex reflection and transmission coefficients for this fabricated material. The experiment performed by Driscoll *et al.* is set up such that both incident and substrate medium are vacuum with $x$ axis parallel to $s$ polarized radiation. In this configuration, the complex reflection coefficients for $p$ and $s$ polarized radiation in Eq. (7) and Eq. (8) reduce to the following:

$$r_{pp} = \frac{\frac{i}{2}\left(\frac{q_{zp}}{k_{z0}\varepsilon_{yy}} - \frac{k_{z0}\varepsilon_{yy}}{q_{zp}}\right)\sin(q_{zp}d)}{\cos(q_{zp}d) - \frac{i}{2}\left(\frac{q_{zp}}{k_{z0}\varepsilon_{yy}} + \frac{k_{z0}\varepsilon_{yy}}{q_{zp}}\right)\sin(q_{zp}d)}$$

$$r_{ss} = \frac{\frac{i}{2}\left(\frac{q_{zs}}{k_{z0}\mu_{yy}} - \frac{k_{z0}\mu_{yy}}{q_{zs}}\right)\sin(q_{zs}d)}{\cos(q_{zs}d) - \frac{i}{2}\left(\frac{q_{zs}}{k_{z0}\mu_{yy}} + \frac{k_{z0}\mu_{yy}}{q_{zs}}\right)\sin(q_{zs}d)}$$

(13)

In Eq. (13), $q_{zp}(\omega)$ and $q_{zs}(\omega)$ have the same definitions as in Eq. (3) and Eq. (4) except for the interchange of the $x$ and $y$ axes to accommodate the experimental set up. $k_{z0}$ is the $z$ component of the free space wave vector. When Eq. (13) is used in conjunction with the reflection fit parameters given in Table 1, it is possible to simulate the $s$ polarization reflectivity across the measured frequency spectrum. Figure 2 illustrates the comparison between the simulated results and the actual experimental results and fitted reflection curves as determined by Driscoll *et al.* [2]. The experimental data and fitted curves over varying AOI are shown in Figure 2 (a). The simulated results are shown in Figure 2 (b) for normal incidence and for an AOI of 40°. As seen in Figure 2, the simulated curves calculated using $4 \times 4$ matrix formalism [Eq. (13)] are identical to the theoretical curves shown in Figure 2 (a). The Driscoll theoretical curve, in turn, is an excellent fit to the measured reflectivity data.

Due to the complexity of the analysis using the Fresnel approach, Driscoll *et al.* [2] constrained themselves to study only the $s$ polarization incident at the sample. $4\times 4$ matrix formalism and full MM measurement should allow more complete analysis of the sample properties using incident light of linear and elliptical polarizations. In order to develop a forward model and analyze the measurements of MMs at oblique angles of incidence, assumptions about the permittivity and permeability along other directions are required. Specifically, assumptions about the $\varepsilon_{yy}(\omega)$ response are necessary in order to illustrate how $4\times 4$ matrix formalism could have been used to predict the MM for this metamaterial. Asymmetries in the SRR fabrication between the $x$ and $y$ axis suggest that $\varepsilon_{yy}(\omega) \neq \varepsilon_{xx}(\omega)$. For purposes of illustration only, we assume that the natural resonance of the $\varepsilon_{yy}(\omega)$ oscillation is 15 GHz as compared to 19.9 GHz for the $\varepsilon_{xx}(\omega)$ oscillation. Eq. (10) is then used to transform the complex reflection coefficients into MM elements. The 8 non-zero elements of the predicted MM applicable to this planar metamaterial are illustrated in Figure 3.

Eq. (10) gives the definitions for each of the MM components in terms of the material's complex reflection coefficients. In Figure 3, the $M_{11}$ component incorporates the modeled reflection intensities of both $p$ and $s$ polarizations. In fact, if $r_{pp}$ were to be mathematically suppressed in this calculation, this MM component would be identical to the reflection graph in Figure 2. The $M_{33}$ component returns the real part of $r_{pp} \times r_{ss}^*$ and, as can be seen, this number is negative throughout the measured frequency spectrum. The $M_{12}$ component returns the difference between the $p$ and $s$ reflected intensities. Figure 3 indicates that this component is close to zero until approximately 11 GHz for the oblique angle case. Finally, $M_{34}$ returns the imaginary part of $r_{pp} \times r_{ss}^*$. The four independent MM components incorporate information about the imaginary parts of the complex reflection coefficients through both the absolute value calculations in $M_{11}$ and $M_{22}$ as well as the product calculations of $M_{33}$ and $M_{34}$.

For proper characterization of materials whose magnetic effects have non-negligible influence on their optical properties, it is important to be able to separate dielectric and magnetic contributions. Spectroscopic experiments usually provide values for the complex refractive index $n = \sqrt{\varepsilon\mu}$ at different frequencies, which do not provide any direct information as to whether it is $\varepsilon$ or $\mu$ which is responsible for a particular feature observed in the spectrum. The difference in the change of the various MM components in response to whether $\varepsilon$ or $\mu$ is changing can separate dielectric and magnetic contributions. For metamaterials, this information is crucial for their design.

This discrimination is indeed possible by performing MM measurements made at varying angles of incidence. To illustrate this point, we model conditions where the index of refraction of a dielectric-magnetic material remains constant but its inputs ($\varepsilon$ and $\mu$) are varied. Specifically, we model a hypothetical case where each of $\varepsilon$ and $\mu$ are allowed to vary between 1 and 6, but their product, $n^2 = \varepsilon\mu$, is held constant at 6. It is evident in Figure 4 (a) that the MM response of the off-diagonal elements is the same in magnitude, but is either positive or negative depending on whether it is $\varepsilon$ or $\mu$ that is changing. The diagonal elements do not change as $n^2$ remains constant. Moreover, as seen in Figure 4 (b), when we introduce the "left handed" [6] material with negative permittivity and permeability, but keeping $\varepsilon\mu = 6$, the $M_{12}$ and $M_{34}$ components respond in opposite directions. This difference in the angular response between $M_{12}$ and $M_{34}$ is an indication of the material being "left handed". This observation is extremely important as it is happening in the thin film sample where the study of such MM measurements at varying AOI may be the only way to identify the anomalous properties of the metamaterial comprising the film. Figure 4 (b) also shows the interesting impedance matching condition discussed in Section 2. When $\varepsilon = \mu$, there is zero reflection at normal incidence.

## 5. Summary

Magnetically active materials in general and metamaterials in particular comprise important classes of materials both from a theoretical perspective as well as for possible device applications. We have presented an analytical approach for the study of these materials using $4\times 4$ matrix formalism. Wave vectors in a dielectric-magnetic medium are derived directly from the eigenvalue solutions of the Berreman equation. We utilized the wave vector approach to derive analytic formulas for the complex reflection and transmission coefficients of thin films whose $\varepsilon$ and $\mu$ tensors match to orthorhombic symmetry. Any other system that has simultaneously diagonalizable $\varepsilon$ and $\mu$ tensors (with coincident principal axes) can be reduced to this case by rotations of the reference frame. We have demonstrated how these calculations can lead directly to the full MM of a non-depolarizing material. Using these results, forward models for the active MM elements of a planar metamaterial were calculated. The separation of the magnetic and dielectric contributions to the optical properties of a material, as well as identification of negative refractive index in a thin film, were demonstrated using measurements of the MM at varying angles of incidence (AOI).

**Acknowledgements**

This work is supported by NSF-MRI: DMR-0821224.

Table 1. Fit parameters for transmission and reflection data [2].

| Parameter | Transmission Fit | Reflection Fit |
|---|---|---|
| $d$ | 0.30 cm | 0.30 cm |
| $A_e$ | 0.7 | 0.7 |
| $\omega_{e0}$ | 22.2 GHz | 19.9 GHz |
| $\omega_p$ | 62 GHz | 62 GHz |
| $\gamma_e$ | 11.6 GHz | 5.4 GHz |
| $A_m$ | .66 | .66 |
| $\omega_m$ | 8.0 GHz | 8.3 GHz |
| $\gamma_m$ | .25 GHz | .19 GHz |

Fig 1. Wave vector diagram of refracted waves propagating in an anisotropic dielectric-magnetic medium.

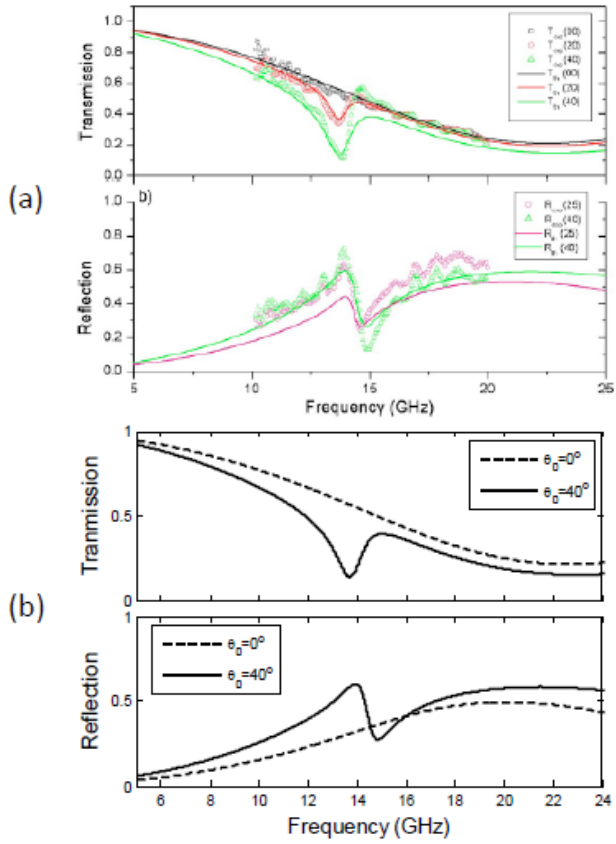

Fig. 2. Reflection and transmission intensities for *s* polarization. (a) Experimental and fitted results determined by Driscoll *et al.* [2]. (b) *s* polarization reflection and transmission reflection coefficient calculated using $4 \times 4$ matrix formalism [Eq. (13)] for two AOI.

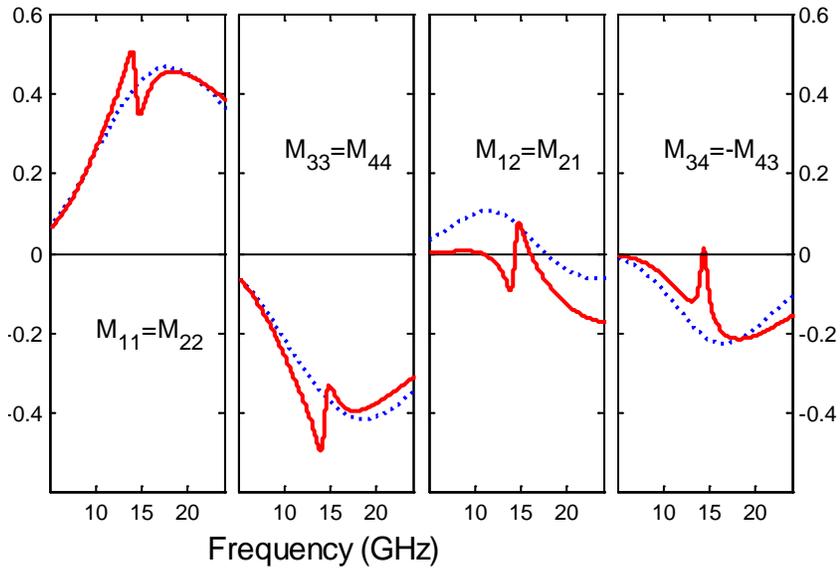

Fig 3. The Mueller Matrix components of a planar metamaterial in the proximity of the resonance at 14 GHz for two AOI. Dotted line $\theta_0 = 0^o$. Solid line $\theta_0 = 40^o$.

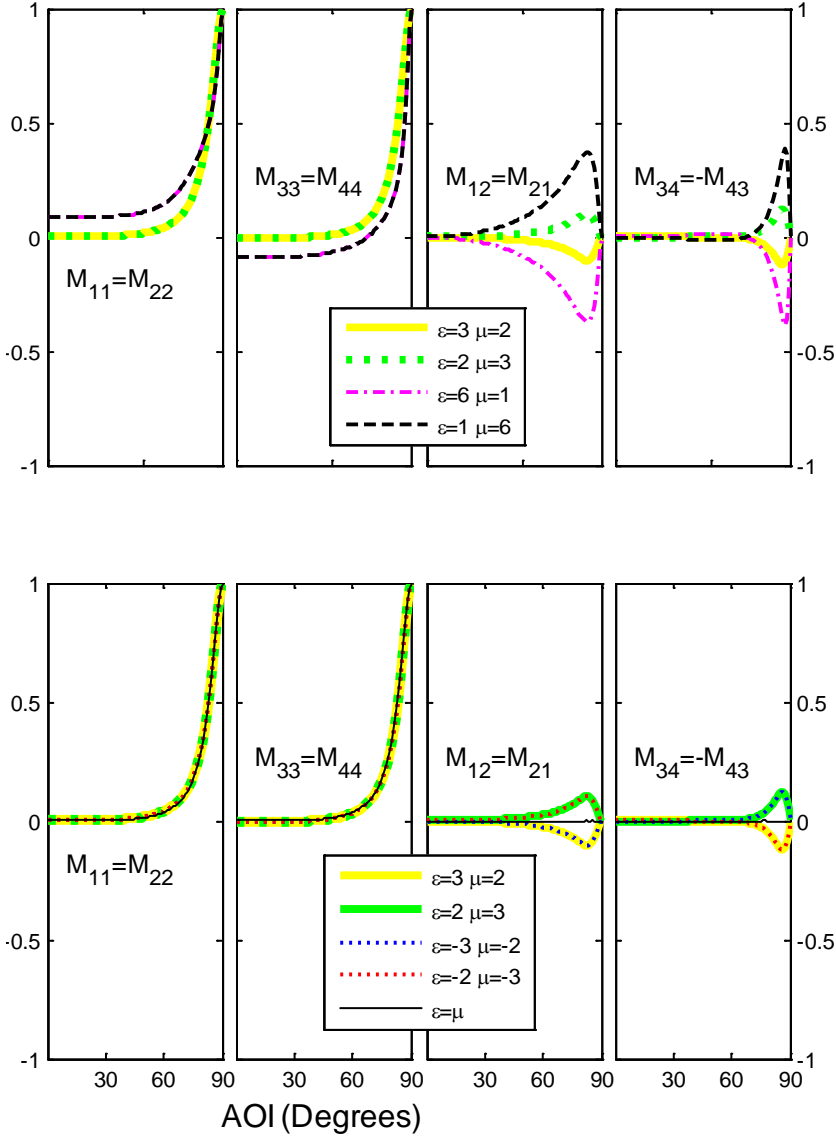

Fig. 4. Distinguishing between magnetic and dielectric contributions in the diagonal and off-diagonal MM components at oblique AOI. (a) Changing $\varepsilon$ and $\mu$ to illustrate the difference in response of $M_{12}$ and $M_{34}$ compared to $M_{11}$ and $M_{33}$. (b) Changing $\varepsilon$ and $\mu$ to illustrate the difference in response of $M_{12}$ compared to $M_{34}$ when "left handedness" is introduced via negative values for $\varepsilon$ and $\mu$.